\documentclass[a4paper,11pt]{article}
\usepackage{pos}
\usepackage{caption}
\usepackage{subcaption}
\usepackage[section]{placeins}
\usepackage{wrapfig}
\usepackage{array,multirow,graphicx}
\newcommand{\eq}[1]{Eq.~(\ref{#1})}

\title{The determination of $r_0$ and $r_1$ in $N_f=2+1$ QCD}

\author*[a]{Tom Matty Bo Asmussen}
\author[a]{Roman Höllwieser}
\author[a]{Francesco Knechtli}
\author[a]{Tomasz Korzec}

\affiliation[a]{Dept. of Physics, Bergische Universität Wuppertal,\\
  Gaußstrasse 20, 42119 Wuppertal, Germany}

\emailAdd{asmussen@uni-wuppertal.de}
\emailAdd{hoellwieser@uni-wuppertal.de}
\emailAdd{knechtli@uni-wuppertal.de}
\emailAdd{korzec@uni-wuppertal.de}

\abstract{We determine the scales $r_0$, $r_1$, the ratio $r_0/r_1$ for $N_f$ = 2 + 1 flavor QCD ensembles generated by CLS. These scales are determined from an improved definition of the static force, which we measure using Wilson loops and furthermore use to study the shape of the potential. Our analysis involves various continuum and chiral extrapolations of data that covers pion masses between 134 MeV and 420 MeV and five lattice spacings down to 0.038 fm.}

\FullConference{The 40th International Symposium on Lattice Field Theory (Lattice 2023)\\
July 31st - August 4th, 2023\\
Fermi National Accelerator Laboratory\\}

\begin{document}
\maketitle

\section{Introduction}
The static potential V(r) plays an important role in lattice QCD and can be computed via Wilson loops. In this work, we use it to determine the scales $r_0$ and $r_1$ which offer a vital link between lattice simulations and physical observables. The calculation has been done using ensambles generated by state-of-the-art simulations of $N_f = 2+1$ flavor QCD by the CLS Consortium \cite{Bruno_2015,Mohler_2018}. We are using HYP-smeared fields to suppress the overlap with excited states, giving a better signal-to-noise ratio. The reduction of systematic errors plays a key role in this analysis as several methods have been used to reduce them to a minimum to give an updated value for the physical value of $r_0$. We will briefly go through how to find the scales from Wilson loops measured on the ensembles. The potential is calculated from those using a GEVP.

\section{Finding the scales through a GEVP}

\begin{figure}
  \centering \includegraphics[height=.48\linewidth]{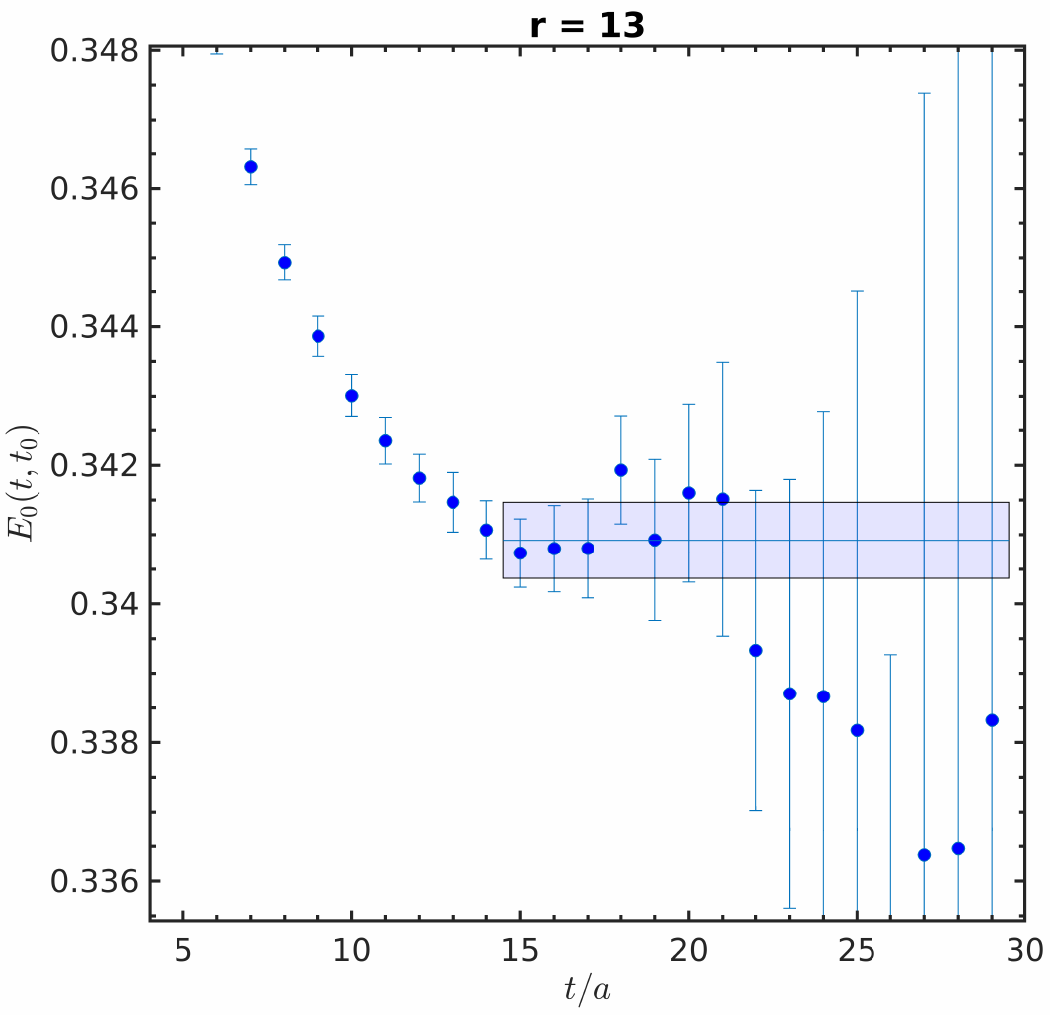}  \includegraphics[height=.48\linewidth]{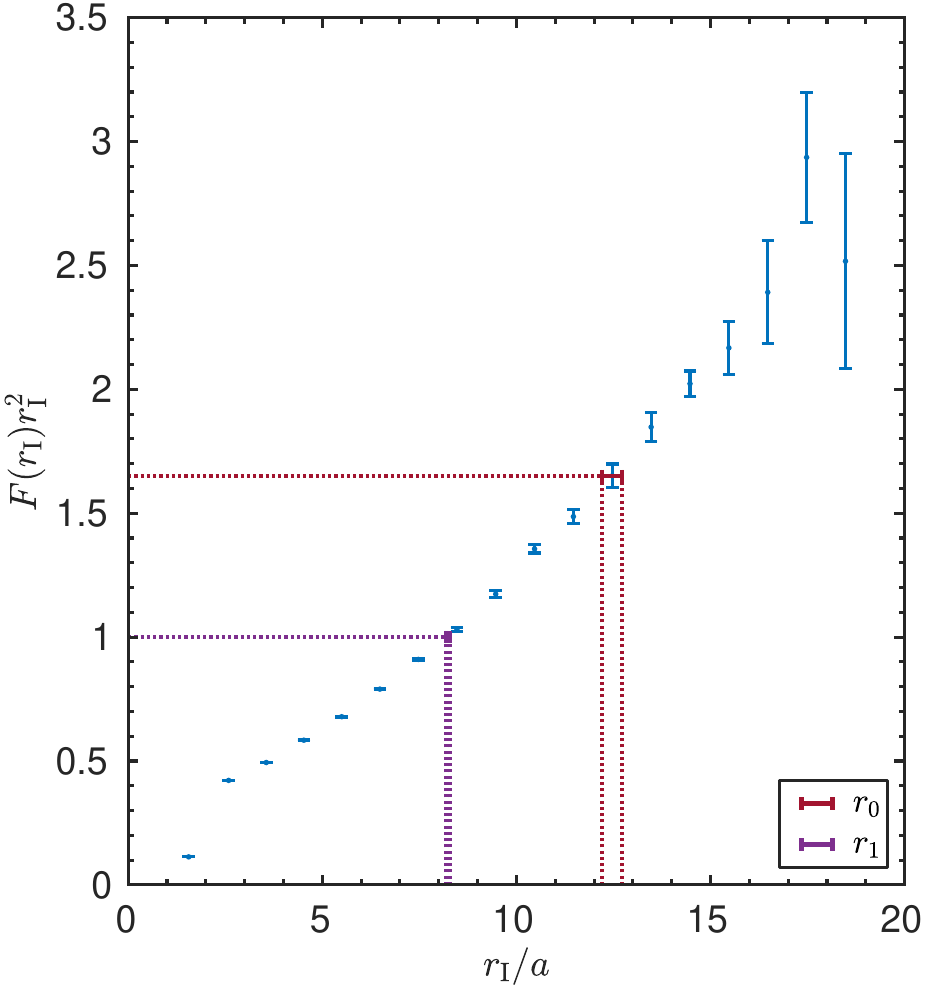}
  \caption{On the left the effective masses for ensemble J501 at r/a = 13 with $t_0$/a = 5 together with the resulting ground state potential calculated from the plateau using a weighted average. On the right the resulting static force together with the $r_1$ and $r_0$ interpolation.}
  \label{fig:GEVP}
\end{figure}

To extract the ground state potential the first step in our method is to solve the generalised eigenvalue problem (GEVP) \cite{Blossier:2009kd} 
on $r/a \times T/a$ on-axis Wilson loops that are measured on the gauge configurations where a correlation matrix is build from different smearing levels \cite{Hasenfratz:2001hp}.
\begin{equation}
    C(t)\;\psi_\alpha = \lambda _\alpha (t,t_G) \;C (t_G)\; \psi_\alpha,
\end{equation}
where $\alpha=0$ corresponds to the ground state and higher values to excited states of the resulting eigenvalues $\lambda _\alpha$. The size of C(t) depends on the amount of smearing levels used. The effective masses are then found by
\begin{equation}
    E_0 (t+\frac{a}{2},t_G) \equiv \ln (\lambda_0 (t,t_G) / \lambda_0 (t+a,t_G)) .
\end{equation}
Performing a weighted average in the plateau region of the effective masses $E_0 (t+\frac{a}{2},t_G)$ as seen in the left plot of figure\ref{fig:GEVP} the ground state potential $V(r)$ for a given r can be calculated. As the excited state contamination is high at early times the start point of the plateau must be chosen with care. To do that, different methods were analysed. The method used in this proceeding was to create a new function that has the starting point of the weighted average as a parameter and to analyse the values and errors given by this function. Furthermore, a fit on the data of the effective mass is performed as a check. The ground state potential can then be used to determine the static force $F(r)$ using the following improved definition
\begin{equation}
    F(r_I) = [V(r)-V(r-a)]/a.
\end{equation}
The improved distance $r_I$ \cite{Sommer:1993ce} is such that the force evaluated at tree level in pertubation theory is $F_{\textrm{tree}}(r_I) = \frac{4}{3} \frac{g_0 ^2}{4\pi r_I ^2}$. It is $r_I = r - a/2 + \mathcal{O}(a^2)$ and suppresses lattice artefacts \cite{Necco:2001xg}. Given the force, we can now find the following scales \cite{Sommer:1993ce,Bernard:2000gd}: 
\begin{align}
    r_I^2 F(r_I) \rvert _{r_I=r_0} &= 1.65 \nonumber\\
    r_I^2 F(r_I) \rvert _{r_I=r_1} &= 1 
    \label{eq:r1&r0}
\end{align}
which will be done by 3 point interpolations using the 3 closest data points. In general, a data point before and two after the scales in \eq{eq:r1&r0} are used for the interpolation. Different amount and ranges have been used as a check. Finally, this gives us the values for $r_0$ and $r_1$ for all ensembles. In the right plot of figure \ref{fig:GEVP} the results of the scales on one ensemble is shown.

\section{Ensembles}

\begin{figure}
    \centering
    \includegraphics[width=.55\linewidth]{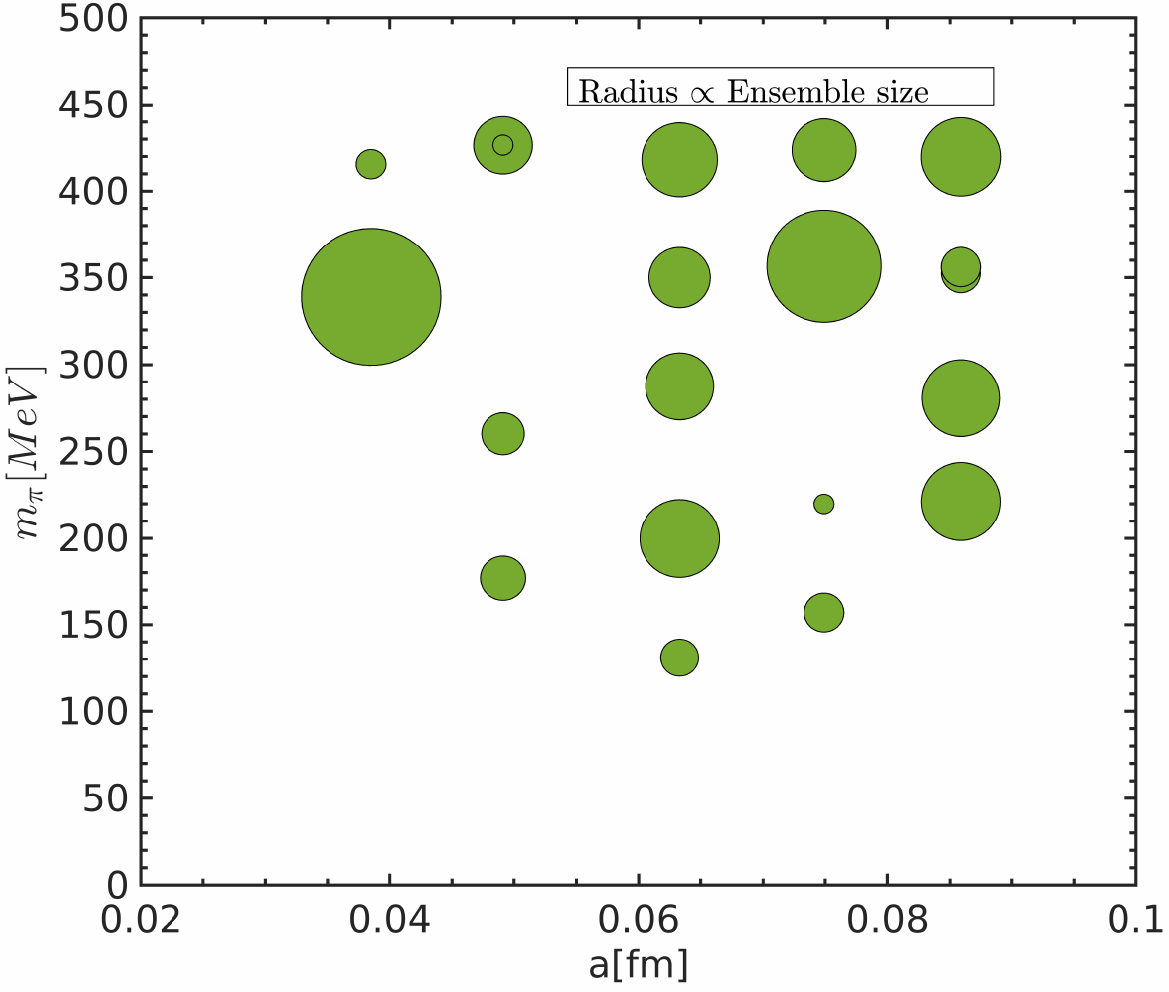}
    \caption{Sketch of the different ensembles with their pion mass versus the lattice spacing. The size of the circles is in relation to the number of configurations spanning 500-3500.}
    \label{fig:circ}
\end{figure}

The Gauge configurations used for the determination of $r_0$ are generated by CLS (Coordinated Lattice Simulations) using $N_f$ = 2 + 1 flavors of nonperturbatively improved Wilson fermions with the Lüscher-Weisz gauge action. 
For most ensembles open boundary conditions in time have been used to address the problem of topological freezing at small lattice spacings and for an improved stability of the simulations a twisted-mass reweighting came into use \cite{Luscher:2012av}. In the analysis we used 20 ensembles, which can be seen in figure \ref{fig:circ} spanning a lattice spacing from 0.085fm to 0.037fm and a pion mass from 430MeV to 134MeV. The statistical error in this work is calculated by the $\Gamma$-method \cite{Wolff:2003sm} where the slow modes in the autocorrelation are explicitly added, as described in \cite{Schaefer:2010hu}.

\section{Results}

\begin{figure}
  \centering
    \includegraphics[height=.49\linewidth]{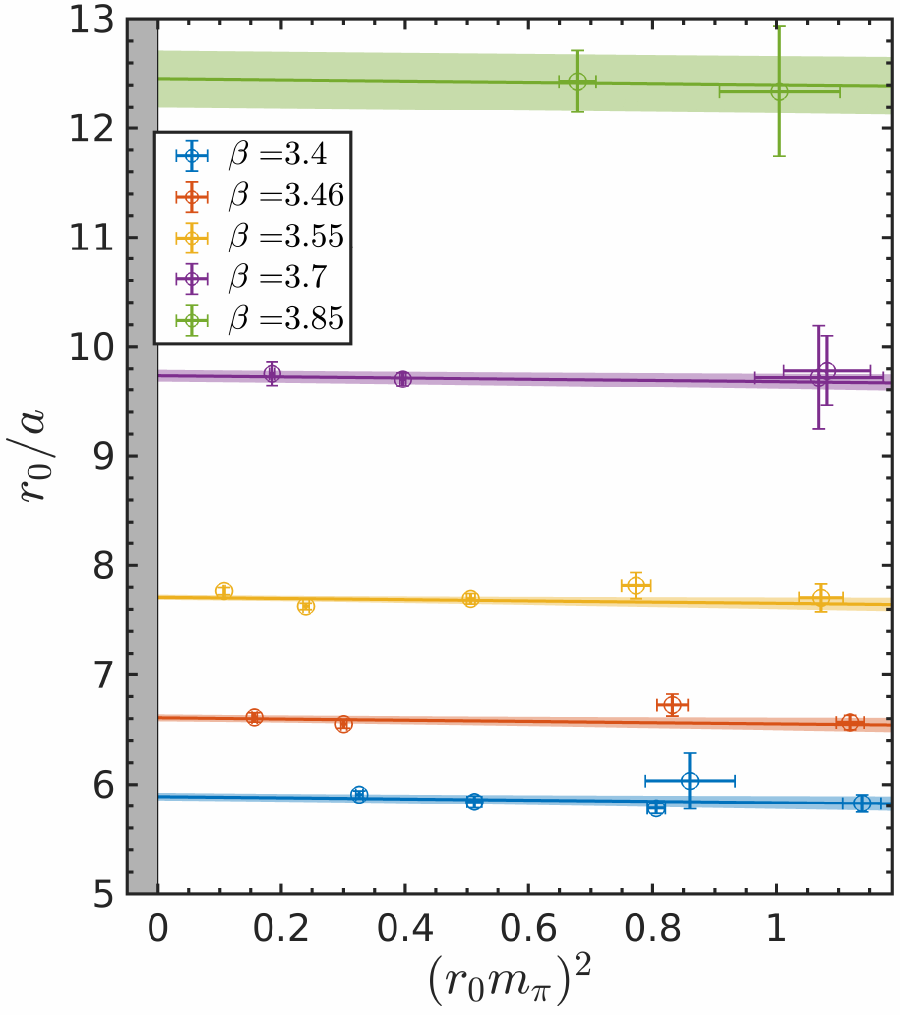}
    \includegraphics[height=.49\linewidth]{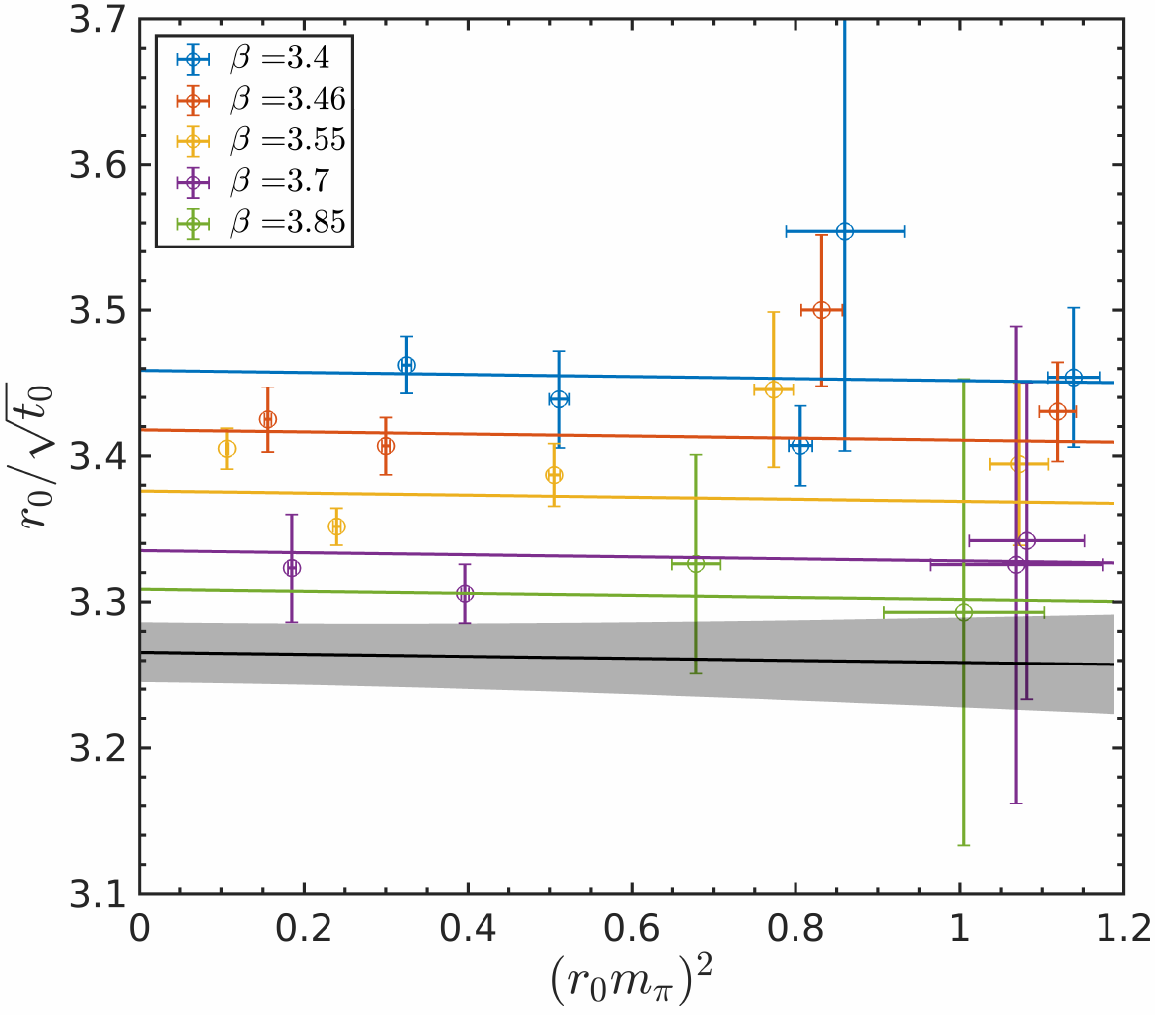}
  \caption{Values of $r_0/a$ for all ensembles in respect to $r_0^2m_\pi^2$ on the left. On the right $r_0/\sqrt{t_0}$ of the different ensembles with a global fit. The grey band corresponds to the continuum extrapolation.}
  \label{fig:r0data}
\end{figure}
\begin{figure}
  \centering
    \includegraphics[height=.45\linewidth]{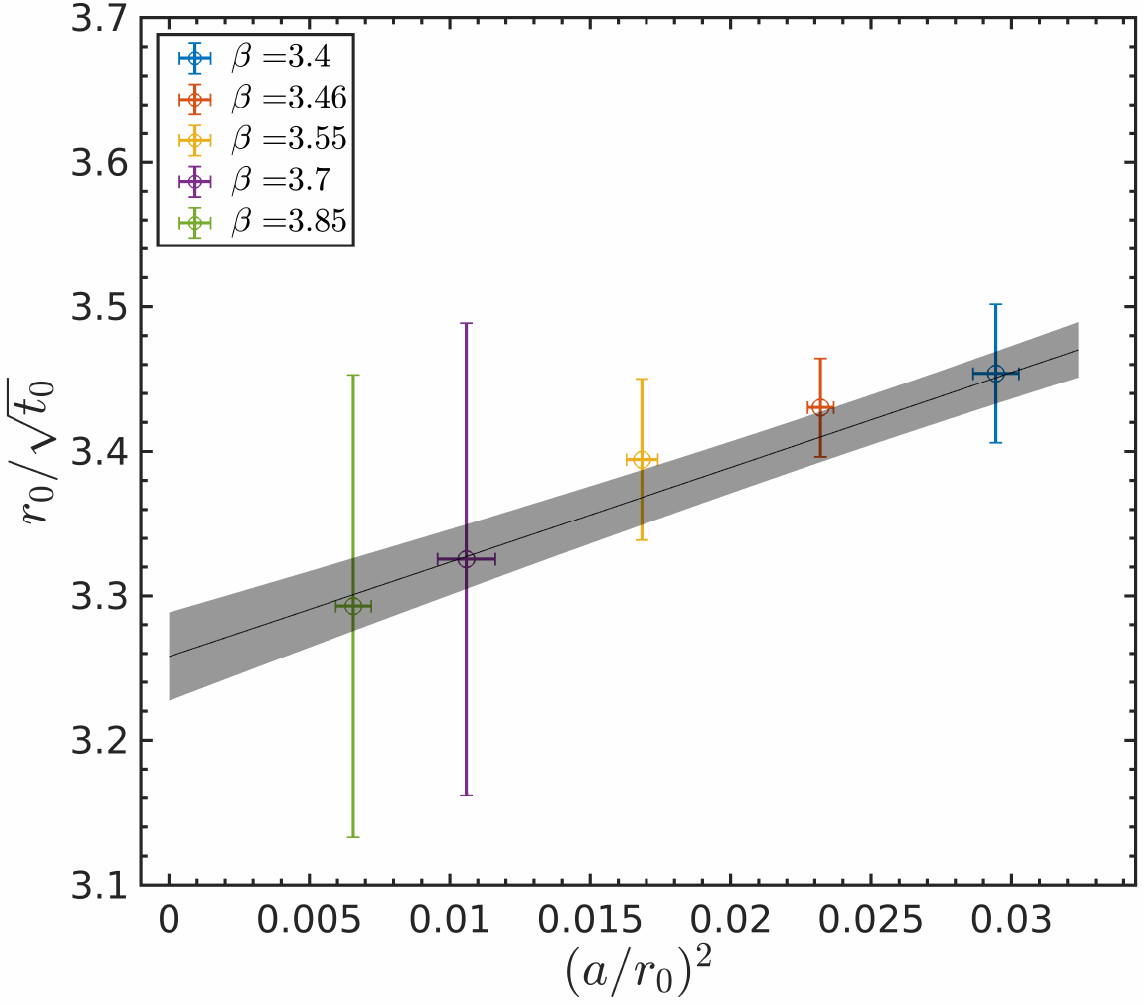}
    \includegraphics[height=.45\linewidth]{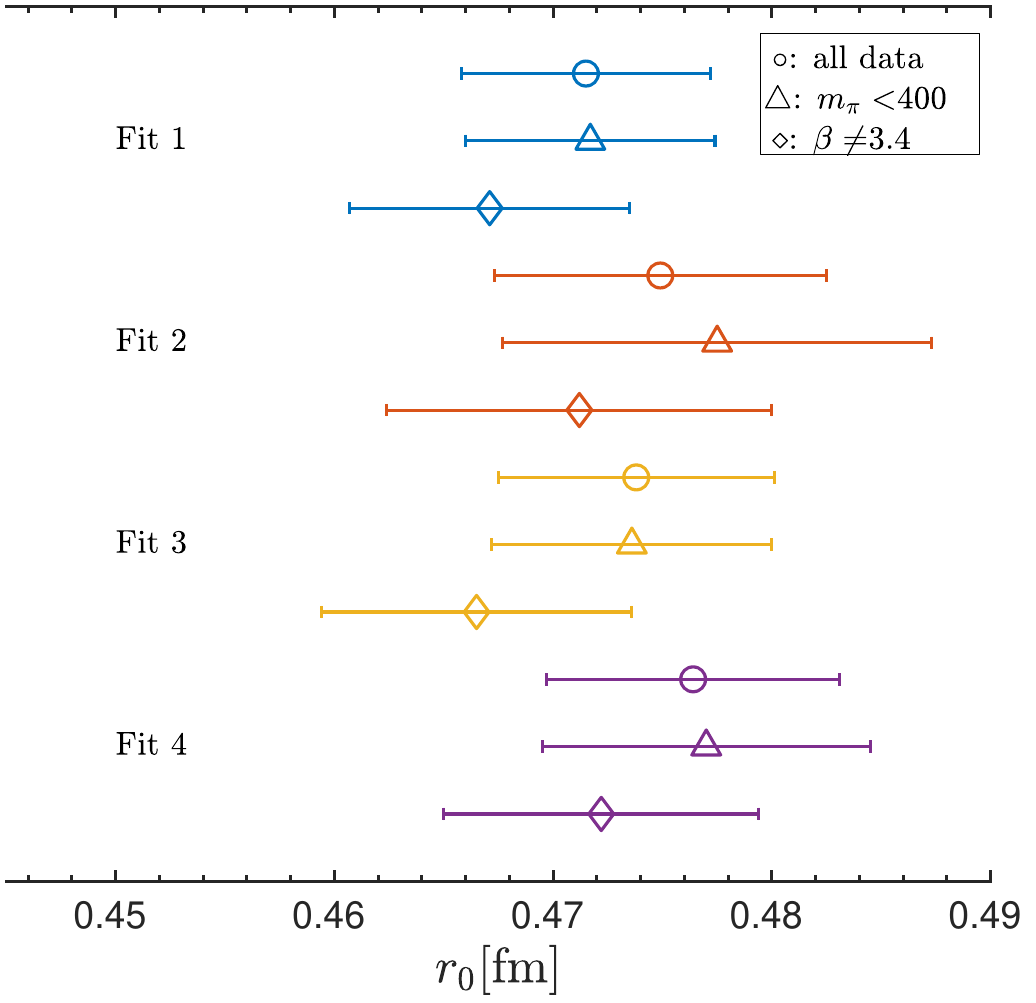}
  \caption{On the left the continuum extrapolation of $\frac{r_0}{\sqrt{t_0}}$ on the ensembles at the symmetric point. On the right the results of $r_0$ at the physical point of the different fits listed in table \ref{table:fits} as a plot.}
  \label{fig:sym&Fit}
\end{figure}

The resulting $r_0$ values are depicted on the left in figure \ref{fig:r0data}. A global fit of the form \begin{equation}
    \frac{r_0}{a} =c_1\rvert_{\beta}  + c_2 (r_0 m_\pi )^2
\end{equation} is performed to the data, which shows a minimal mass-dependence of $r_0$. 
To find the physical value of $r_0$ we extrapolate the data into the continuum and (inter\footnote{One ensemble is generated at the physical pion mass})extrapolate to the physical mass of $m_\pi$ with help of the physical scale $\sqrt{t_0}=0.1443(7)$fm \cite{Strassberger:2021tsu} using the global fit: 
\begin{equation}
    \frac{r_0}{\sqrt{t_0}}=c_1 + c_2 \left( \frac{a}{r_{0,\textrm{sym}}}\right) ^2 + c_3 (r_0 m_\pi )^2.
    \label{eq:Fig-Fit}
\end{equation}
$r_{0,\textrm{sym}}$ is calculated at the symmetric point where the quark masses are degenerate. The grey band in figure \ref{fig:r0data} depicts the mass dependence in the continuum limit. As a small check of the fit-parameters, we used the fit parameters from the global fit from \eq{eq:Fig-Fit} and compared the results of the ensembles with with symmetric masses by evaluating the fit at the symmetric point, which can be seen on the left in figure \ref{fig:sym&Fit}, where the fit aligns with the symmetric points. In total, 4 different fits were used where \eq{eq:Fig-Fit} is labeled as Fit 1. Fit 2 has an added mass term to add a mass dependence on the lattice artifact. Fit 3 is constructed in a way to change $c_3$ to be independent of $r_0$, but instead indirectly by $t_0$ using $\phi_2 = 8t_0 m_\pi ^2$ and $\phi_4 = 8t_0 \left( m_K ^2 + \frac{1}{2}m_\pi ^2\right)$ which are used to shift the masses, since the simulation conditions of the ensembles to have a constant sum of the bare quark masses might not correspond to constant $\phi_4=1.098$\cite{Strassberger:2021tsu} due to for instance discretization effects. The Fit 4 is done by having the parameters only depend on $t_0$ in contrast to $r_0$. All the fits have also been performed using cuts with respect to high un-physical pion masses and the coarsest lattices which are on table \ref{table:fits} with the resulting physical values of $r_0$ and $\chi^2 / d.o.f.$. On the right of figure \ref{fig:sym&Fit} the results of the fits are visualized. The results of the different fits agree with each other, with the largest change obtained by excluding the coarsest lattices, which can be due to the loss of a large portion of data points. Fit 3 shows that there is not a large discrepancy by neglecting a mass shift using the mass derivatives of $\phi_4$\cite{Bruno_2017}. We will use the results of Fit 1 for our final result which  together with the results from different groups published in the FLAG Review 2021\cite{Aoki_2022} is shown in figure \ref{fig:Flag}.

\begin{table}[!]
\centering
    \begin{tabular}{ c| c c  | c c}
        \hline
         Method &$r_0[\textrm{fm}]$ & $\chi ^2$ /d.o.f. &$r_0[\textrm{fm}]$ & $\chi ^2$ /d.o.f.\\
          \hline
          \multicolumn{3}{c|}{F1: $c_1+c_2 (\frac{a}{r_{0,sym}})^2 + c_3(r_0 m_\pi)^2$ } &\multicolumn{2}{c}{F3:  $c_1+c_2 (\frac{a}{r_{0,sym}})^2 + c_3 \phi_2 + c_4(1.098-\phi_4)$}\\
        \hline
         all & 0.4715(57) & 19.5/17 & 0.4738(63) & 17.1/16\\
         $m_\pi<$400 & 0.4717(57) & 18.3/11 & 0.4736(64) & 16.8/10\\
         $\beta\neq$3.4 & 0.4671(64) & 14.2/12  & 0.4665(71) & 9.1/11  \\
        
        \hline
          \multicolumn{3}{c|}{F2:  $c_1+c_2 (\frac{a}{r_{0,sym}})^2 + c_3(r_0 m_\pi)^2 + c_4(m_\pi a)^2$ } & \multicolumn{2}{c}{F4:  $c_1+c_2 (\frac{a}{r_{0,sym}}) + c_3(t_0 m_\pi^2) $  }\\
          \hline
        all & 0.4749(76) & 11.8/16 & 0.4764(67) & 17.4/16\\
        $m_\pi<$400 & 0.4775(98) & 9.1/10 & 0.4770(75) & 15.6/10\\
        $\beta\neq$3.4 & 0.4712(88) &  9.8/11 & 0.4722(72) & 12.4/11\\
          \hline
    \end{tabular}
    \caption{A table of the different fit parameters that have been tested together with their $\chi ^2$ /d.o.f. and results of the physical $r_0$.}
    \label{table:fits}
\end{table}

\subsection{$r_1$ and $r_0 / r_1$}

\begin{figure}
  \centering
    \includegraphics[height=.41\linewidth]{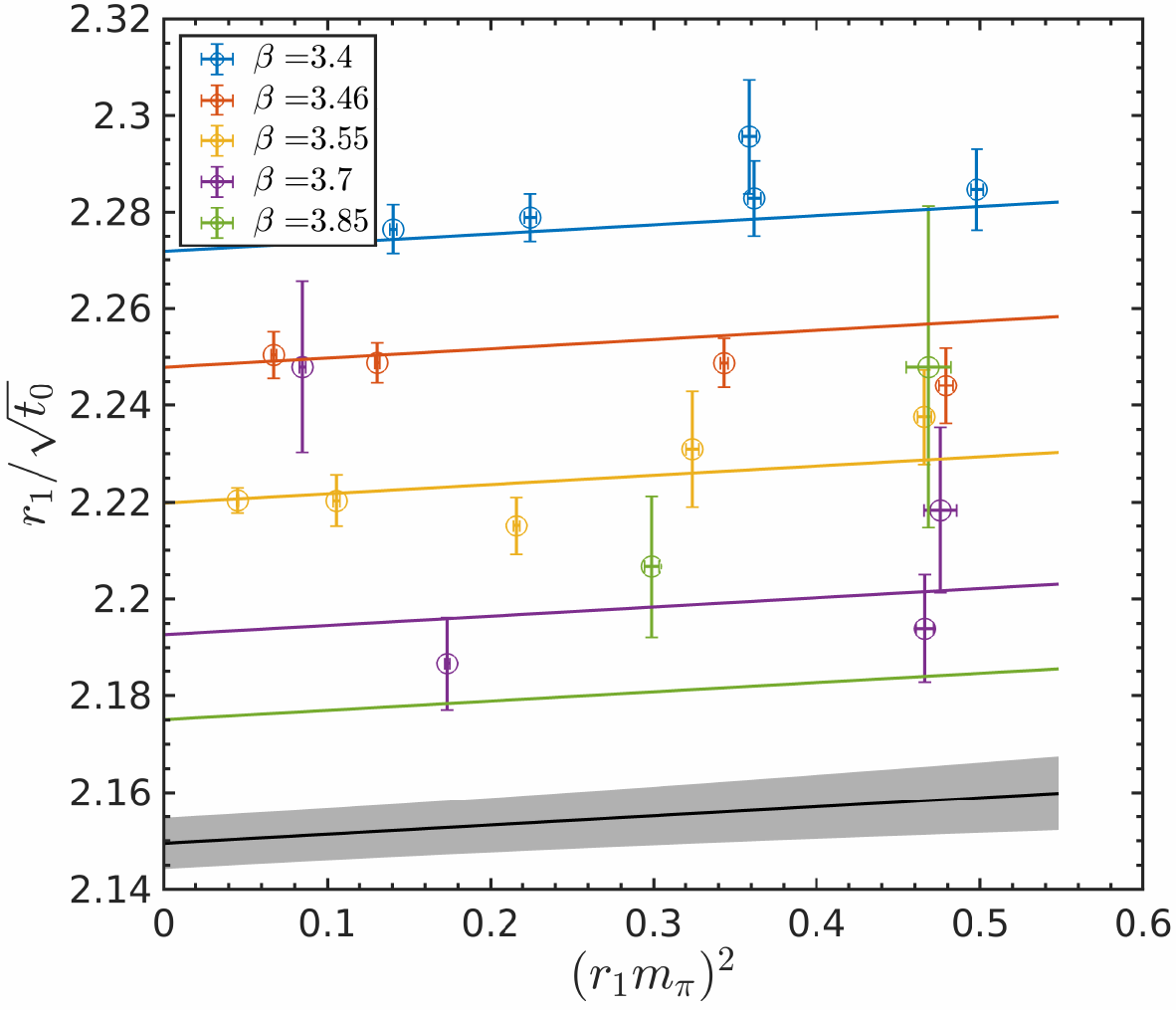}
    \includegraphics[height=.41\linewidth]{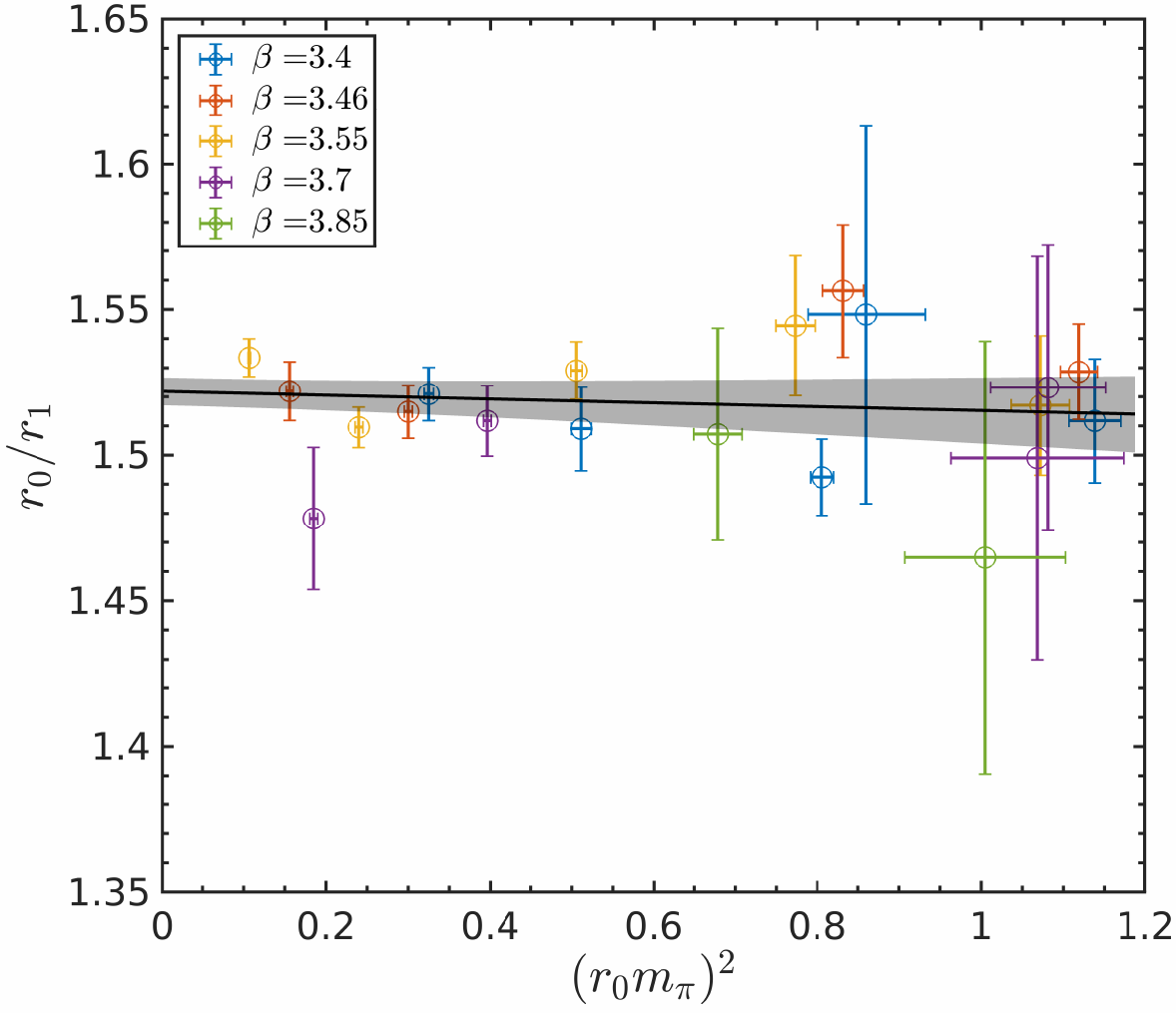}
  \caption{On the Left $r_1/\sqrt{t_0}$ of the different ensembles with a global fit. The grey band corresponds to the continuum extrapolation.  On the right $r_0/r_1$ of the different ensembles with a linear fit given as the grey band.}
  \label{fig:r1&r1r0}
\end{figure}

A similar analysis as for $r_0$ has been done for the scale $r_1$ where the data points with the continuum extrapolation using a global fit are on the left of figure \ref{fig:r1&r1r0}. What can be seen here is that several ensembles of the finer lattice spacings are more than one sigma away from the results of the global fit for their corresponding lattice spacing. In general the determination of $r_1$ has been troublesome in some cases for the finer lattice spacings (see for example the green points in the left plot of figure \ref{fig:r1&r1r0}). On the right plot of the same figure \ref{fig:r1&r1r0} the ratio $r_0/r_1$ is shown together with the result of a linear fit as the gray band, giving a nearly constant result. Some outliers, in particular the value corresponding to $\beta=3.7$ with the pion mass closest to the physical value, will be analyzed further. The physical values of the result $r_1$ and the ratio $r_0/r_1$ compared to the other results of the FLAG Review 2021\cite{Aoki_2022} are shown in figure \ref{fig:Flag}.

\subsection{Shape of the potential}

\begin{figure}[h!]
    \centering
    \includegraphics[height=.5\linewidth]{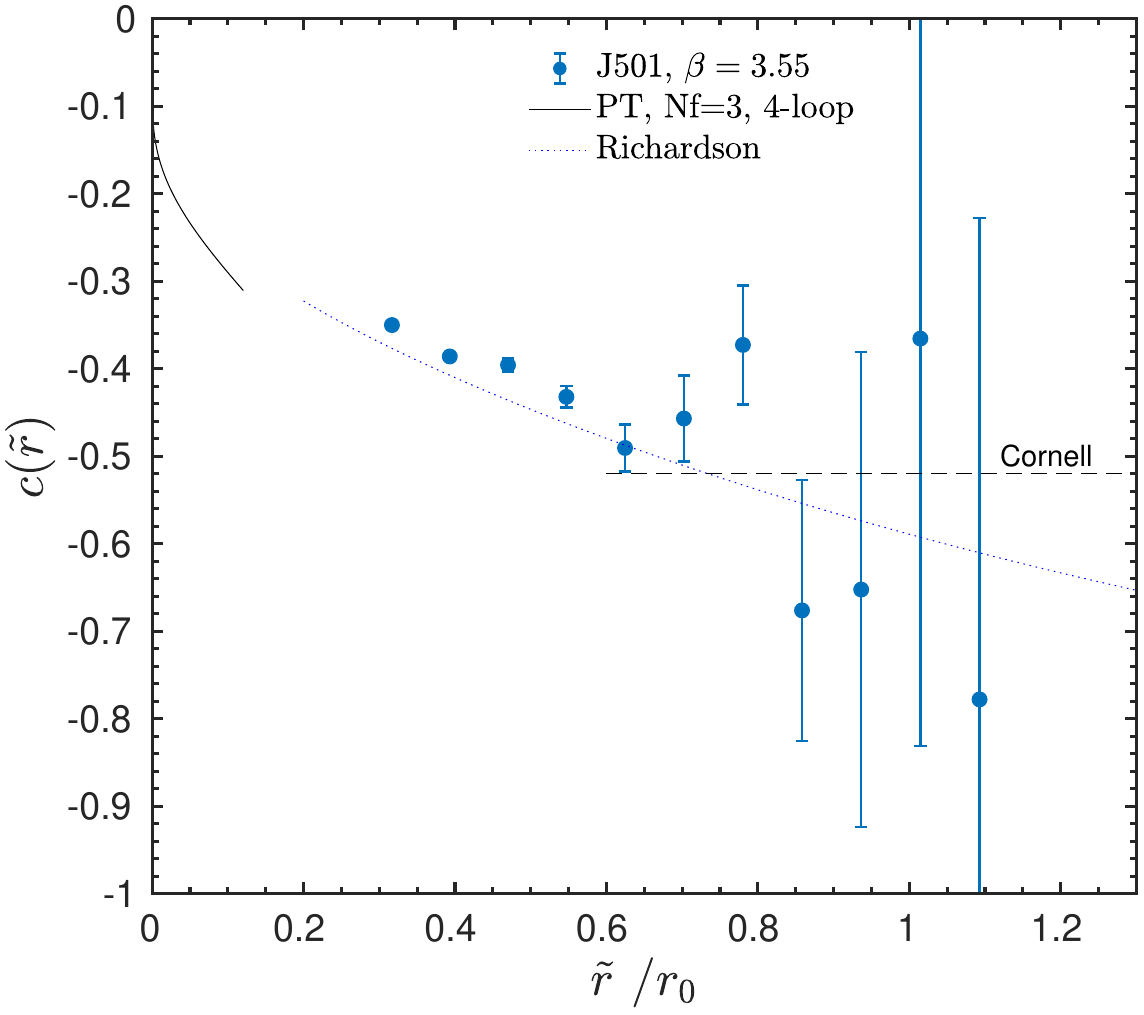}
    \captionof{figure}{The physical quantity $c(r)$ from \eq{eq:Shape}  for the finest lattice spacing with the smallest available pion mass compared with the pertubation theory at short distances. For comparison we plot the curve for the Richardson potential\cite{RICHARDSON1979272} and the Cornell value $c=-0.52$ \cite{PhysRevD.21.203}. }
    \label{fig:shape}
\end{figure}

We studied the shape of the potential by building $c(r)= \frac{1}{2} r^3 F'(r)$ computed as:

\begin{align}
    c(\Tilde{r}) =\frac{1}{2} \Tilde{r}^3 [V(r+a)+V(r-a)-2V(r)]/a^2
    \label{eq:Shape}
\end{align}
If the potential would be a pure Cornell potential $V=\sigma r -\frac{K}{r}$ then $c=-K$. The result of the quantity $c(\Tilde{r})$ for one of the fine ensembles is presented in figure \ref{fig:shape}. $\Tilde{r}$ in this case is an improved distance analogous to $r_I$ defined in \cite{Luscher:2002qv}. For short distances, $c(r)$ behaves as prescribed by the perturbation theory, which for comparison is plotted for $N_f=3$ using a 4-loop beta function for the c-scheme which can be found in \cite{Donnellan:2010mx}. For comparison the Richardson potential\cite{RICHARDSON1979272} is plotted as well, which the data follows quite closely. One can see that the data gets quite noisy around and after $\Tilde{r}=r_0$. It will be analyzed further together with other ensembles.

\section{Conclusion}

\begin{figure}[h]
  \centering
    \includegraphics[height=.3\linewidth]{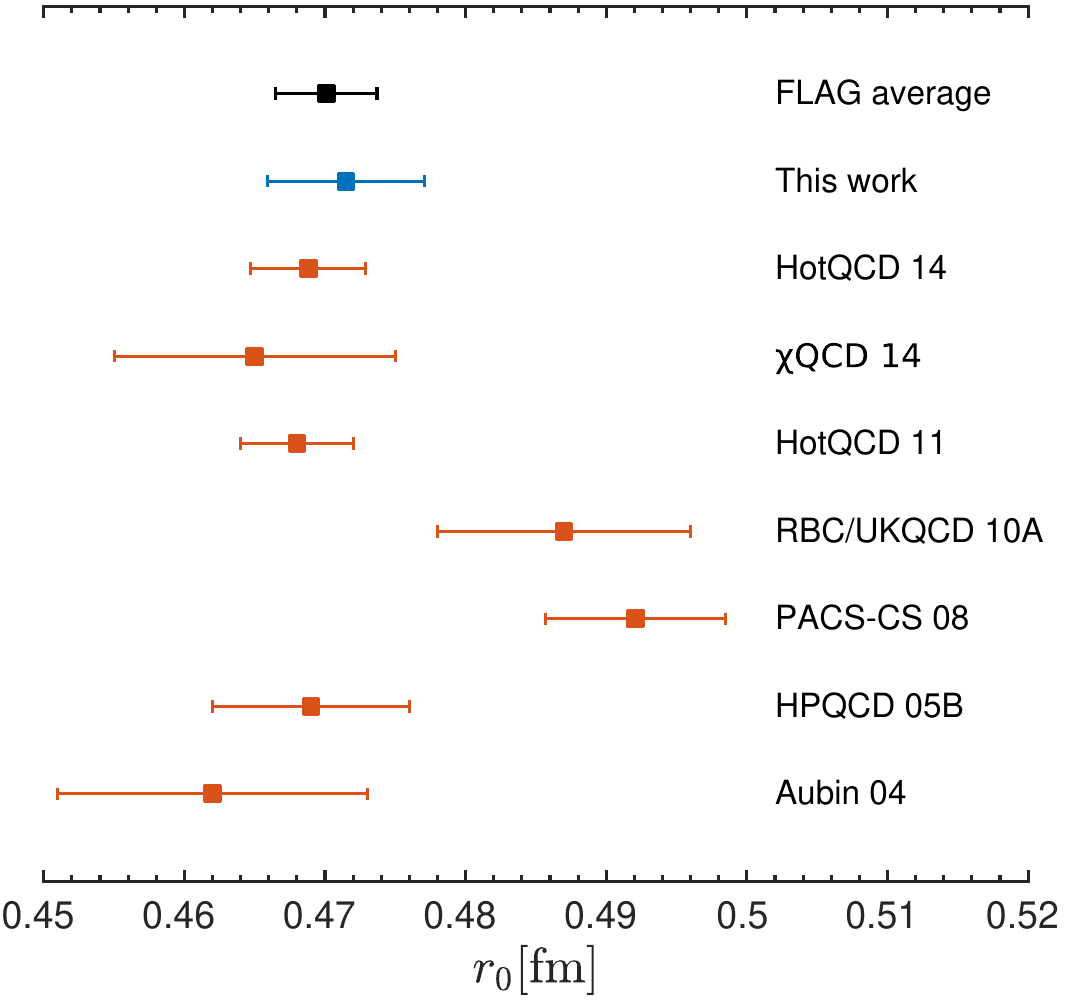}
    \includegraphics[height=.3\linewidth]{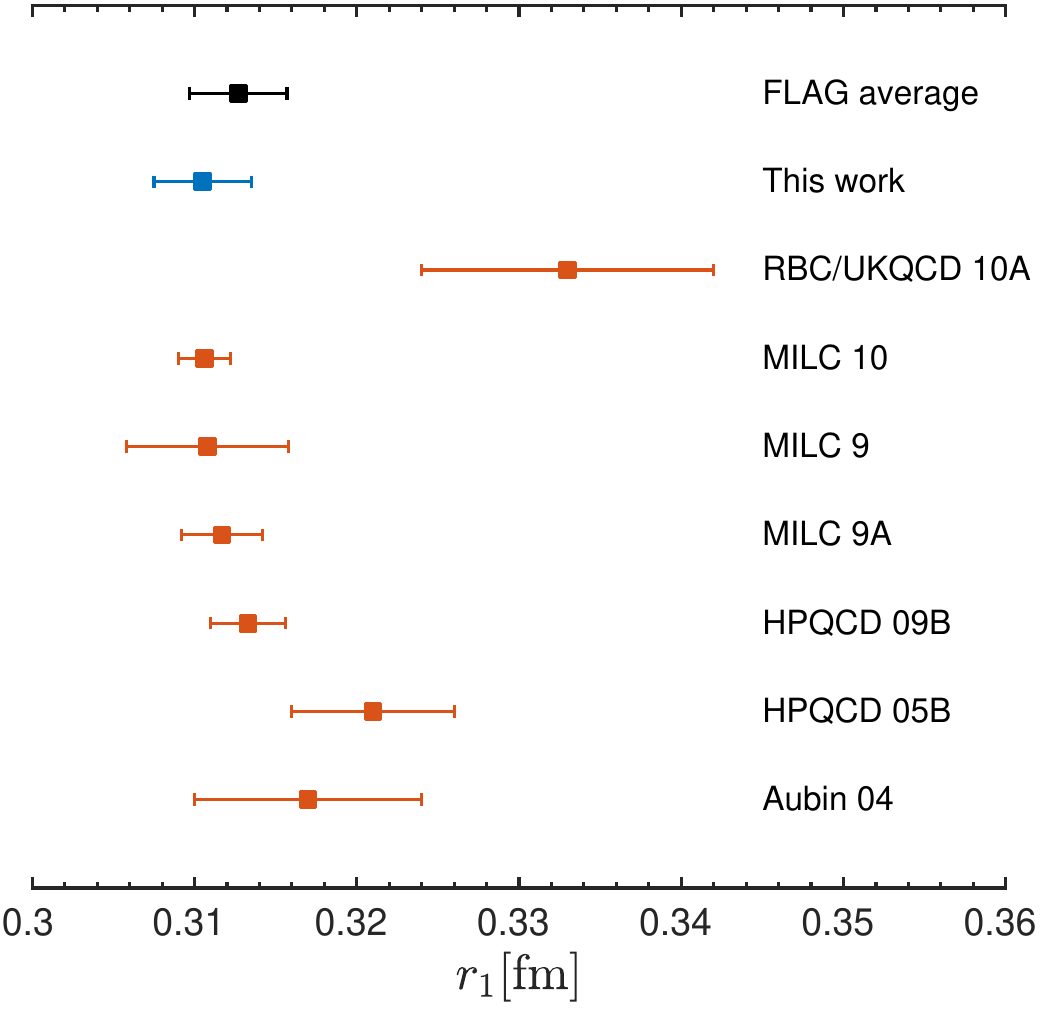}
   \includegraphics[height=.3\linewidth]{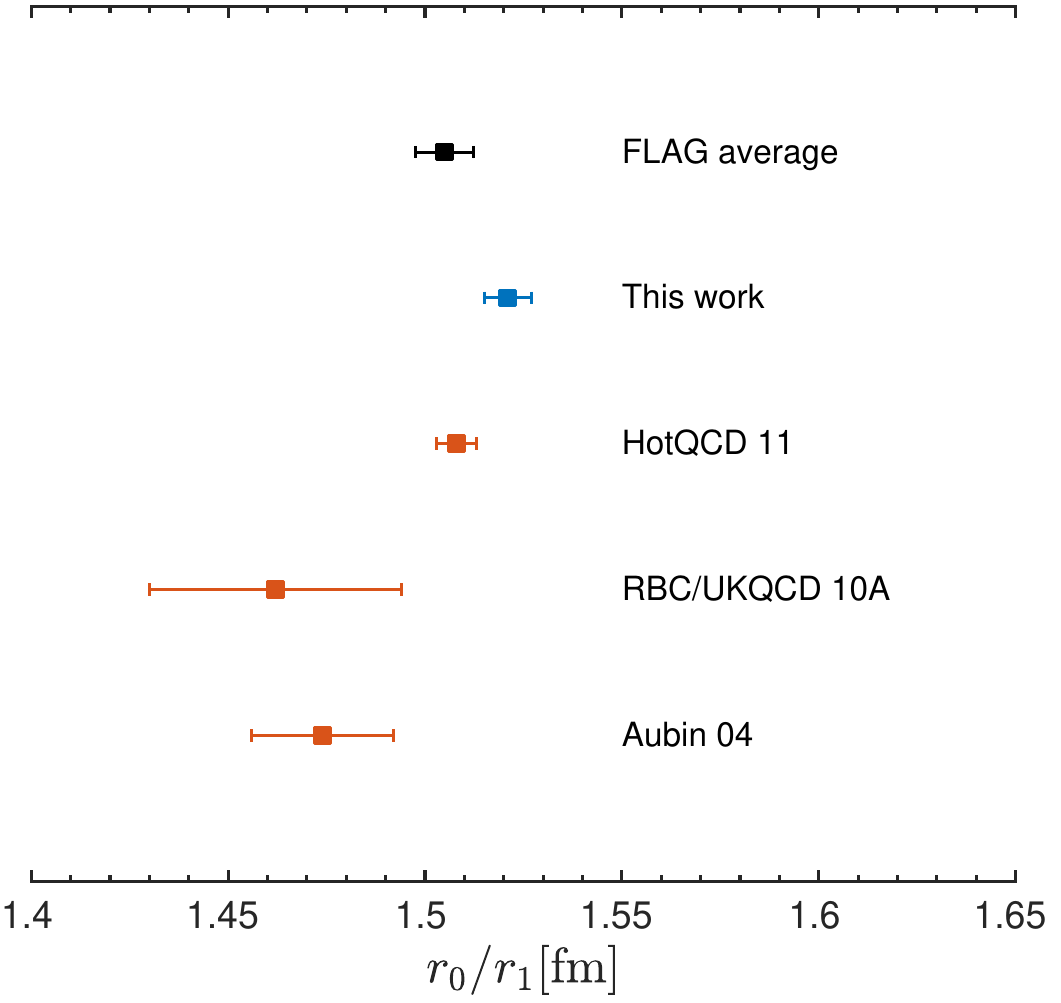}
  \caption{Comparison between other results presented in Flag21\cite{Aoki_2022} and our results (blue points) of $r_0$(left), $r_1$(middle), and $r_0/r_1$(right) at the physical point.}
  \label{fig:Flag}
\end{figure}

In this work, we have presented a new detailed analysis to find the values of $r_0$ and $r_1$. A comparison to other groups from the Flag report of 2021\cite{Aoki_2022} is presented in figure \ref{fig:Flag}. Several methods have been used to control systematic error, which will be discussed extensively in detail in an upcoming article. We conclude that for finer lattices, $r_0$ can be extracted more reliably than $r_1$. 

\section{Acknowledgements} The authors gratefully acknowledge the Gauss Centre for Supercomputing e.V. (www.gauss-centre.eu) for funding this project by providing computing time on the GCS Supercomputer SuperMUC-NG at Leibniz Supercomputing Centre (www.lrz.de). Some computations were carried out on the PLEIADES cluster at the University of Wuppertal, which was supported by the Deutsche Forschungsgemeinschaft (DFG) and the Bundesministerium für Bildung und Forschung (BMBF). The work is supported by the German Research Foundation (DFG) research unit FOR5269 "Future methods for studying confined gluons in QCD". The project is receiving funding from the programme " Netzwerke 2021", an initiative of the Ministry of Culture and Science of the State of Northrhine Westphalia, in the NRW-FAIR network, funding code NW21-024-A (R.H.). The sole responsibility for the content of this publication lies with the authors. We thank Rainer Sommer for valuable discussions.

\bibliographystyle{utphys} 
\bibliography{paper}


\end{document}